\documentclass[aps,prl,twocolumn,showpacs,floatfix,superscriptaddress]{revtex4-1}
\usepackage{color,graphicx}
\usepackage{soul}
\usepackage{hyperref}
\usepackage{natbib}
\begin{document}
\flushbottom

\title{Vortex creep at very low temperatures in single crystals of the extreme type-II superconductor Rh$_9$In$_4$S$_4$}

\author{Edwin~Herrera}
\affiliation{Laboratorio de Bajas Temperaturas y Altos Campos Magn\'eticos, Unidad Asociada UAM, CSIC, Departamento de F\'isica de la Materia Condensada, Instituto de Ciencia de Materiales Nicol\'as Cabrera and Condensed Matter Physics Center (IFIMAC), Universidad Aut\'onoma de Madrid, Spain}

\author{Jos\'e~Benito-Llorens}
\affiliation{Laboratorio de Bajas Temperaturas y Altos Campos Magn\'eticos, Unidad Asociada UAM, CSIC, Departamento de F\'isica de la Materia Condensada, Instituto de Ciencia de Materiales Nicol\'as Cabrera and Condensed Matter Physics Center (IFIMAC), Universidad Aut\'onoma de Madrid, Spain}

\author{Udhara~S.~Kaluarachchi}
\affiliation{Ames Laboratory, Ames, IA 50011}
\affiliation{Department of Physics $\&$ Astronomy, Iowa State University, Ames, IA 50011}

\author{Sergey~L.~Bud'ko}
\affiliation{Ames Laboratory, Ames, IA 50011}
\affiliation{Department of Physics $\&$ Astronomy, Iowa State University, Ames, IA 50011}

\author{Paul~C.~Canfield}
\affiliation{Ames Laboratory, Ames, IA 50011}
\affiliation{Department of Physics $\&$ Astronomy, Iowa State University, Ames, IA 50011}

\author{Isabel~Guillam\'on}
\affiliation{Laboratorio de Bajas Temperaturas y Altos Campos Magn\'eticos, Unidad Asociada UAM, CSIC, Departamento de F\'isica de la Materia Condensada, Instituto de Ciencia de Materiales Nicol\'as Cabrera and Condensed Matter Physics Center (IFIMAC), Universidad Aut\'onoma de Madrid, Spain}

\author{Hermann~Suderow}
\affiliation{Laboratorio de Bajas Temperaturas y Altos Campos Magn\'eticos, Unidad Asociada UAM, CSIC, Departamento de F\'isica de la Materia Condensada, Instituto de Ciencia de Materiales Nicol\'as Cabrera and Condensed Matter Physics Center (IFIMAC), Universidad Aut\'onoma de Madrid, Spain}

\date{\today}

\begin{abstract}
We image vortex creep at very low temperatures using Scanning Tunneling Microscopy (STM) in the superconductor Rh$_9$In$_4$S$_4$ ($T_c$=2.25 K). We measure the superconducting gap of Rh$_9$In$_4$S$_4$, finding $\Delta\approx 0.33$meV and image a hexagonal vortex lattice up to close to H$_{c2}$, observing slow vortex creep at temperatures as low as 150 mK. We estimate thermal and quantum barriers for vortex motion and show that thermal fluctuations likely cause vortex creep, in spite of being at temperatures $T/T_c<0.1$. We study creeping vortex lattices by making images during long times and show that the vortex lattice remains hexagonal during creep with vortices moving along one of the high symmetry axis of the vortex lattice. Furthermore, the creep velocity changes with the scanning window suggesting that creep depends on the local arrangements of pinning centers. Vortices fluctuate on small scale erratic paths, indicating that the vortex lattice makes jumps trying different arrangements during its travel along the main direction for creep. The images provide a visual account of how vortex lattice motion maintains hexagonal order, while showing dynamic properties characteristic of a glass. 
\end{abstract}

\maketitle
\section{Introduction}
\vspace{-0.4cm}
In type-II superconductors, vortex-vortex repulsion favors an ordered vortex lattice. This competes with vortex pinning and thermal fluctuations that favor disordered vortex arrangements\cite{Blatter94,Brandt95}. Under the action of a cu\-rrent, vortices move and the superconductor leaves the zero resistance state. Vortex motion producing a re\-si\-dual dissipation in absence of an applied current has been observed in many materials and is termed vortex creep. Vortex creep has awakened interest of experiment and theory alike for a long time, because it ultimately limits achieving a true dissipationless state in a superconductor.

Vortex creep can occur in any practical situation in a superconducting specimen. For example, when changing the magnetic field below the superconducting critical temperature $T_c$ in a type-II superconductor, vortices enter from the edges of the sample jumping over pinning barriers and filling the interior\cite{Blatter94,Brandt95,Larkin79,RevModPhys.76.471}. The magnetization usually acquires a near equilibrium situation after some time, which is mostly quite short. However, often the magnetization continues to vary over much longer times due to vortex creep. Vortex creep is also observed when field cooling from above $T_c$, because strongly pinned metastable vortex lattices are created when crossing the peak effect region\cite{PhysRevLett.77.2077,Paltiel00}. Vortex creep is driven by a current density $j$, which results from the difference between the actual magnetization from the metastable vortex configurations and its equilibrium value. $j$ is well below the critical current density $j_c$\cite{Feigelman89}.

A large Ginzburg-Levanyuk number $G_L$ \cite{Blatter94,Brandt95,Feigelman89,Kes89,PhysRevLett.60.2202,Blatter93,PhysRevLett.85.4948,PhysRevB.86.024515,Eley16,Levanyuk59,Ginzburg60} favors vortex creep. $G_L$ is the ratio between the critical temperature and the superconducting condensation energy ($\propto H_c^2 \xi^3$, with $H_c$ the thermodynamic critical field) in a volume of size of the coherence length $\xi$ and is given by $G_L=\frac{1}{2}\left(\frac{\mu_o k_BT_c}{4 \pi H_c^2 \xi^3}\right)^2$. $G_L$ quantifies the relevance of thermal fluctuations in a superconducting material and ranges from $\approx 10^{-1}$ in cuprate superconductors to $\approx 10^{-10}$ in conventional low $T_c$ superconductors.

Vortex creep has been studied thoroughly using macroscopic techniques, such as magnetization or resistivity (see e.g. Refs.\cite{PhysRevLett.81.3231,PhysRevB.65.144519,PhysRevB.78.224506,PhysRevB.86.024515,PhysRevB.43.8709,Klein14,Eley16}). However, imaging experiments at low temperatures are, to our knowledge, scarce. 

Previous imaging studies mostly address vortex flow driven by a metastable magnetic field configuration within the peak effect regime or at high temperatures\cite{Pardo98,Troyanovski99,Troyanovski02,Guillamon11b,Lee11}. Moving vortex lattices have been imaged in real space at very low magnetic fields using magnetic decoration and at high magnetic fields using STM\cite{Pardo98,Troyanovski99,Troyanovski02,Guillamon11b,Lee11}. They have shown different structure factors changing from hexagonal to smectic-like \cite{Olson98}. Smectic structures forming channels where vortex positions are uncorrelated along the flow direction but correlated perpendicular to it appear more often at low magnetic field where intervortex interaction is weak \cite{Moon96}. In contrast, crystalline-like hexagonal moving lattices are predicted for stiff dense lattices under high driving currents suggesting that motion can induce order in the vortex lattice. The latter, known as moving Bragg glass, are free from topological defects and show long range positional and orientational correlations \cite{Giamarchi94,Giamarchi95}. For instance, the smectic driven phase has been observed in 2H-NbSe$_2$ at extremely low fields and high driving currents, with crystalline order appearing at higher fields \cite{Pardo98}. In our experiment, contrary to previous cases, the flow velocity is low and we measure at  high fields in the dense lattice regime and at very low temperatures where the stiffness is maximum, so that motion is expected to be correlated.

Furthermore, in previous experiments, vortex motion occurs due to currents of order of $j_c$. Here we directly observe slow vortex creep at millikelvin temperatures for a situation where $j<<j_c$. We find that in this regime vortices move by jumps with large directional changes and show strong fluctuations on top of the overall motion in the direction for creep.

We have used single crystals of the recently discovered superconducting material Rh$_9$In$_4$S$_4$ ($T_c$=2.25 K) \cite{Udhara16}. The low temperature coherence length is of $\xi\approx 9.4$ nm. The penetration depth is very large, of nearly 600 nm and the mean free path is rather low, of approximately 5 nm (see Appendix). The Ginzburg-Landau parameter is of $\kappa\approx 61$, showing that this material is an extreme type-II superconductor. The Ginzburg-Levanyuk parameter is of $G_L\approx 1.78 \times 10^{-5}$. This value is above the one found usually in low $T_c$ superconductors and shows that thermal effects are important for the vortex lattice of Rh$_9$In$_4$S$_4$. The peak effect is observed in a magnetic field range similar to that found in many other strongly disordered superconducting single crystalline materials \cite{Udhara16}. As we show below, vortex creep is particularly apparent in this compound, probably due to the combination of a large peak effect region and a relatively large Ginzburg-Levanyuk parameter.

First we show basic characterization of this material from tunneling density of states and vortex imaging. We find a superconducting tunneling conductance at 150 mK that is spatially homogeneous at zero magnetic field with superconducting gap  $\Delta=0.33$ meV. We also find that the vortex lattice is hexagonal at low temperatures. Then we image creeping vortex lattices and determine the main parameters of vortex motion (direction and velocity) directly from our images by following each vortex during experiments lasting for many hours.

\vspace{-0.4cm}
\section{Experimental Methods}
\vspace{-0.3cm}
Single crystals of Rh$_9$In$_4$S$_4$ were grown using a solution growth technique \cite{Canfield92,Canfield01} from high-purity Rh, In and S elements. Details of the synthesis and sample characterization are described in Ref. \cite{Udhara16}. We used crystals with lateral sizes between 1-3 mm and transversal areas less than 1 mm$^2$ perpendicular to [100]. Figure \ref{Fig1} shows a picture of the samples. To prepare the surface, we tried to cleave the samples in both ambient and cryogenic conditions using the pulling system described in Ref.\cite{Suderow11}. We combine this with either a ceramic blade or a piece of brass glued on top of the crystals so as to break the crystals by pushing on the slab at low temperatures. Most often crystals did not break and the few ones that broke revealed irregular surfaces which were not shiny and where we did not achieve stable tunneling conditions as we find usually in metals with a cleaving plane. We obtained however excellent results by using the pristine as-grown samples. Scanning conditions at low temperatures were reproducible and independent of the tunneling conductance.

To obtain the tunneling conductance, we take the numerical derivatie of the I-V curve, as in previous work\cite{Suderow11,Crespo06a,Herrera15}. The magnetic field is applied parallel to the $c$-axis. To obtain the vortex lattice images shown in this work, we map the zero bias conductance normalized to its value at high bias. Neither filtering nor image treatment is made to the images we present here.

\vspace{-0.4cm}
\section{Tunneling conductance and vortex lattice of {Rh$_9$In$_4$S$_4$}}
\vspace{-0.3cm}
\begin{figure}[h!]
\includegraphics[width=0.48\textwidth]{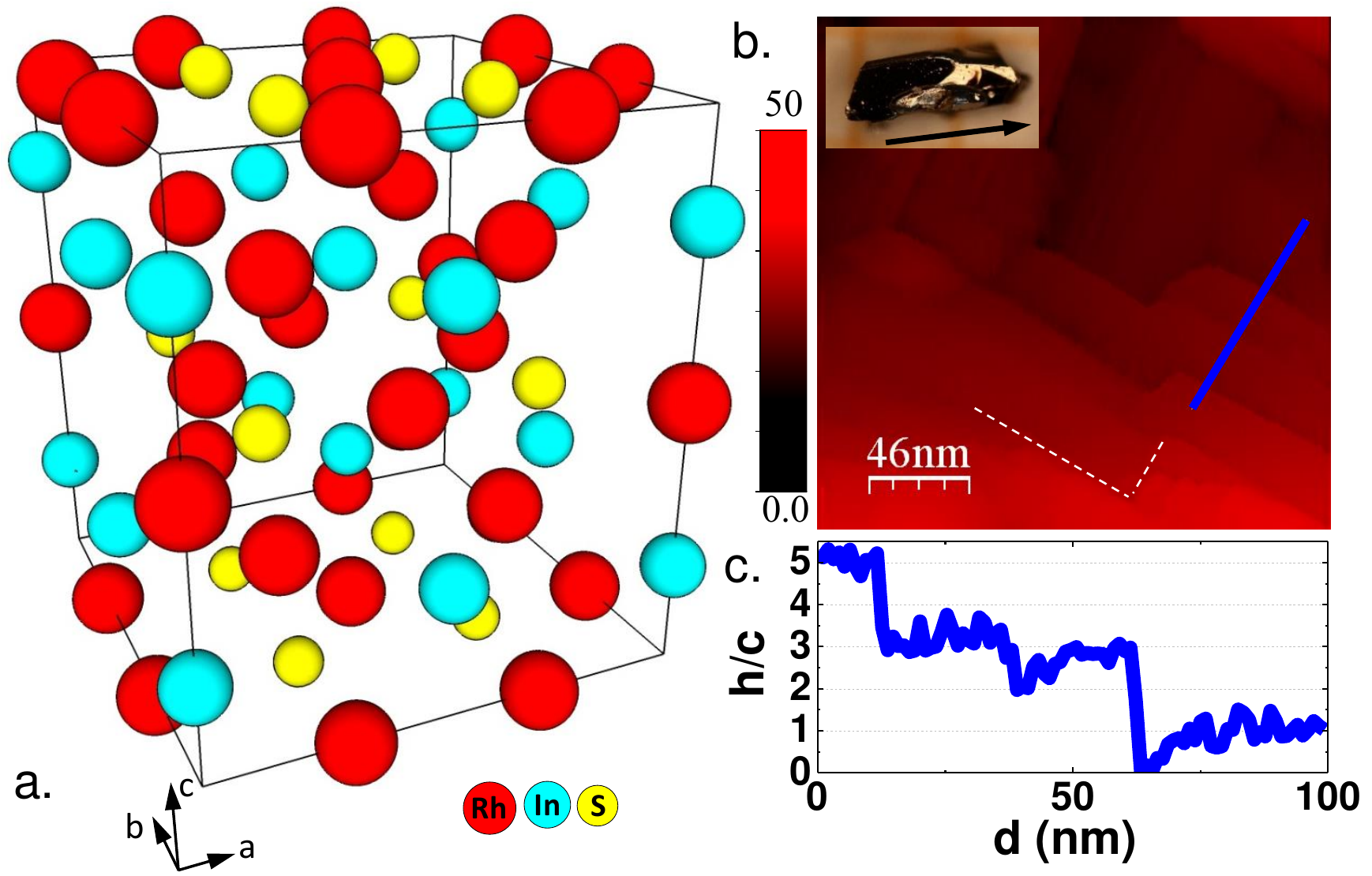}
\caption{(Color online) (a) Unit cell of Rh$_9$In$_4$S$_4$ with lattice parameters $a$ = 7.7953(3) \r{A}, c = 8.8583(3) \r{A} (see Ref.\cite{Udhara16} for more details). (b) STM topographic image obtained at 150 mK. The color bar gives the contrast in the image in \AA. The inset shows a photograph of a single crystal on a mm grid, with the arrow pointing along the [100] direction. The magnetic fields were applied perpendicular to [100] direction. The white dashed line represent the right angle at the corners of the steps. (c) Profile along the blue line in the top right panel. We normalized the vertical displacement to the {\it c-axis} parameter of Rh$_9$In$_4$S$_4$.}
\label{Fig1}
\end{figure}

Fig.\ref{Fig1}a shows the crystal structure of Rh$_9$In$_4$S$_4$. In the topographic STM image shown there (Fig.\ref{Fig1}b), we can find square symmetric features in the image, separated by steps (Fig.\ref{Fig1}c, in the appendix we provide further topographic images). The step height is an integer multiple of the $c$-axis unit cell, pointing out that the observed intertwined squares and rectangles are due to the growth procedure that can be expected in a single crystalline material with tetragonal symmetry.

\begin{figure}[h!]
\includegraphics[width=0.48\textwidth]{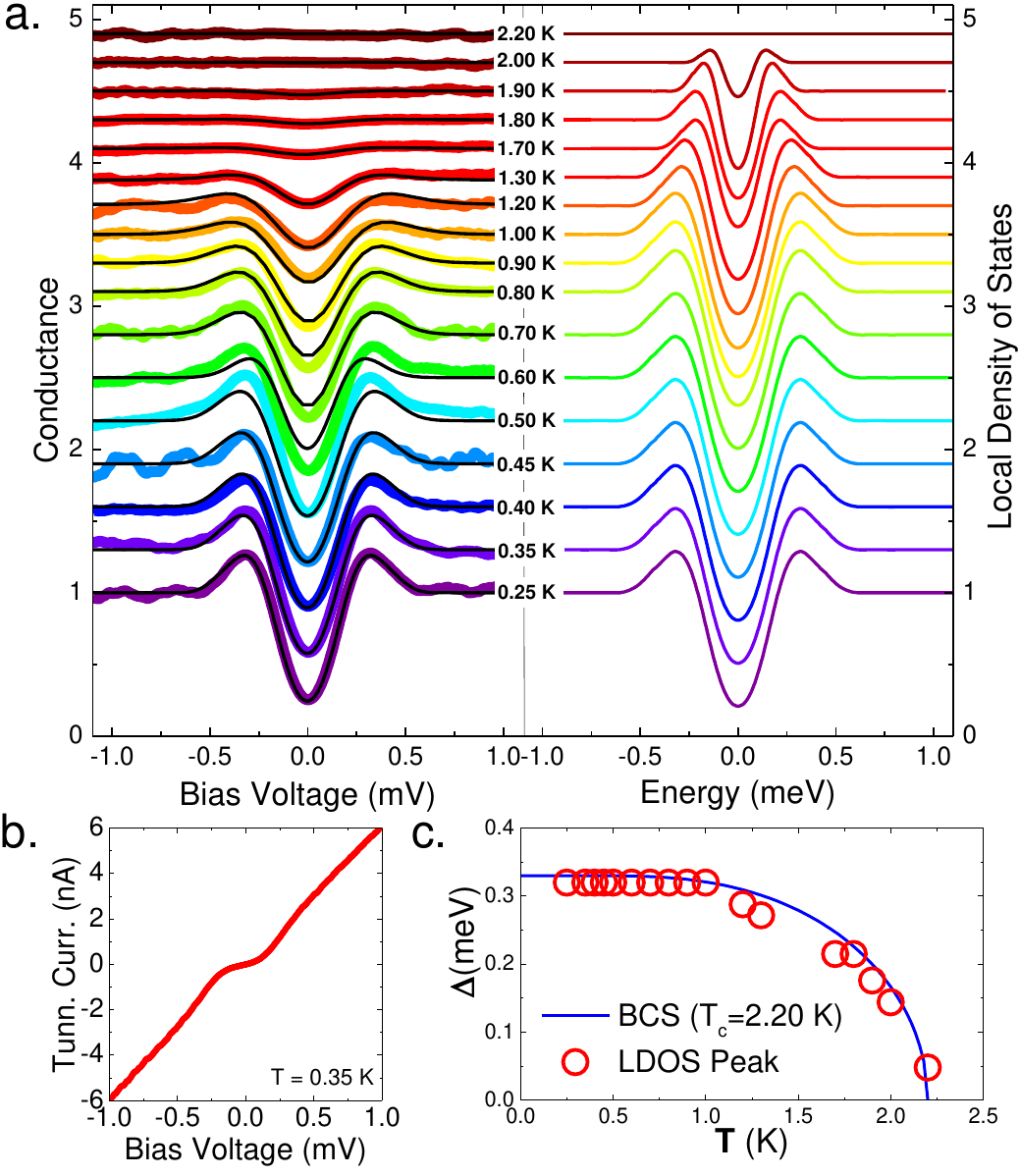}
\vskip -0.2cm
\caption{(Color online) (a) (Left panel) Tunneling conductance curves vs bias voltage as a function of  temperature. Black lines are the fits obtained from convoluting the density of states vs energy curves shown in the right panel with the derivative of the Fermi function at each temperature. (b) IV curve at 0.35 K. (c) Temperature dependence of the position in energy of the quasiparticle peaks of the density of states (right panel in (a)). The  blue line is the expression from BCS theory, using $\Delta$=1.76$ k_BT_c$=0.33 meV obtained with $T_c$=2.3 K.
}
\label{Fig2}
\end{figure}

The tunneling conductance as a function of bias voltage for different temperatures is shown in Fig.\ref{Fig2}a (left panel). At the lowest temperatures we observe a smeared tunneling conductance that deviates strongly from the high quasiparticle peaks and zero bias conductance expected in a standard BCS s-wave superconductor (see e.g.\cite{Herrera15}). The tunneling conductance remains mostly unaltered as a function of the position over the surface. In order to find the superconducting gap, we deconvolute the density of states from the tunneling conductance, see Fig.\ref{Fig2}a (right panel) (using the derivative of the Fermi function at each temperature, see e.g. Ref.\cite{Crespo06a}). We observe that the density of states becomes featureless at a temperature of 2.3 K, which coincides with the bulk transition temperature $T_c$\cite{Udhara16}. The quasiparticle peaks in the density of states are located at an energy of 0.33 meV, which is the value expected for the superconducting gap within simple s-wave BCS theory ($\Delta$=$1.76k_BT_c$). The temperature dependence of the position of the quasiparticle peaks also follows expectations from BCS theory (Fig.\ref{Fig2}c). 

The size of the superconducting gap found with our measurements is somewhat below the one extracted from macroscopic specific heat measurements. These lead to a slightly enhanced at the superconducting transition observed in specific heat, pointing out possible strong coupling effects\cite{Udhara16}. Our measurements show rather smeared density of states, which points out that there can be a distribution of values of the superconducting gap with larger values than the one obtained from the above analysis. More striking is however the presence of a finite density of states at the Fermi level. The bulk density of states is clearly zero within the gap, as shown by specific heat measurements\cite{Udhara16}. Other intermetallic compounds with surfaces prepared in similar conditions show zero tunneling conductance within the gap\cite{Rubio01,Herrera15}. Some sort of surface contamination could eventually lead to a normal contribution to the tunneling conductance by, for instance, a normal surface layer or pair breaking. This being said, it is quite remarkable that we find the bulk $T_c$ and $H_{c2}$, as well as the value of $\Delta$ expected within s-wave BCS theory from our tunneling experiments. 

\begin{figure*}
\includegraphics[width=1\textwidth]{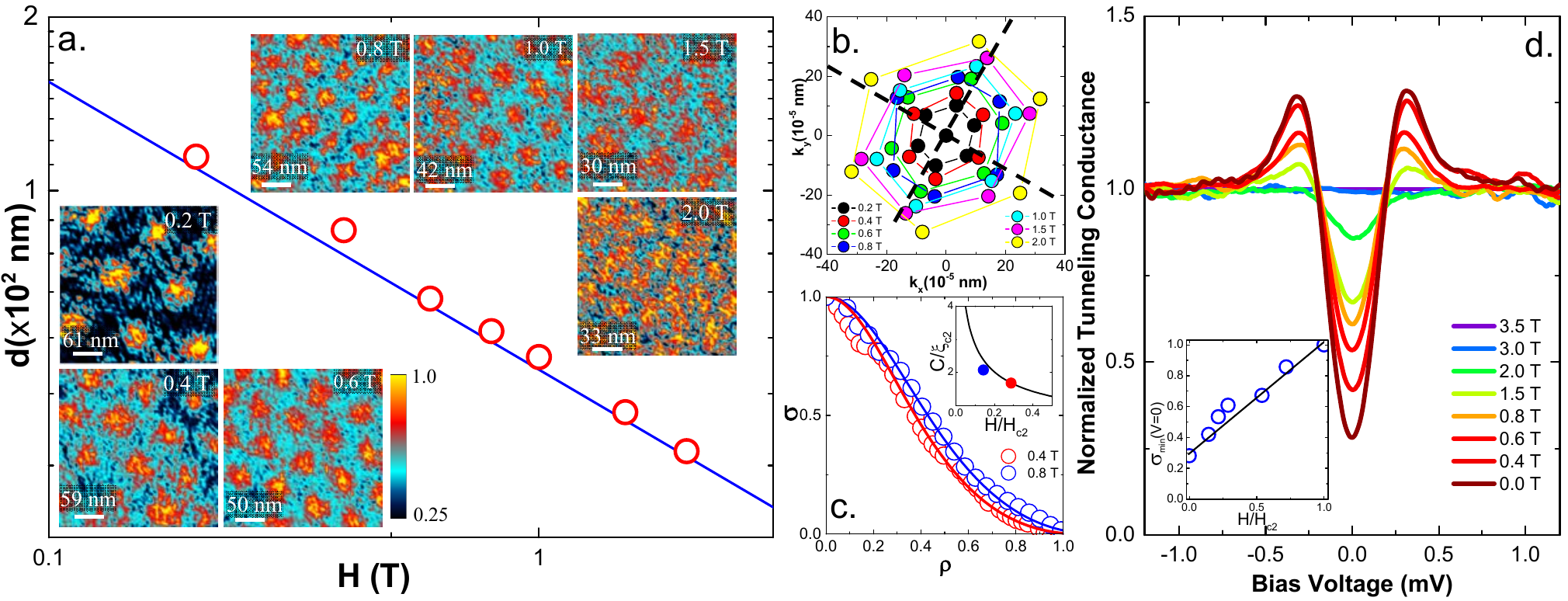}
\vskip -0.2cm
\caption{(Color online) {\bf (a)} Vortex lattice evolution as a function of the magnetic field at 150 mK. The insets show zero bias conductance images at different magnetic fields (and at different locations). Contrast in the zero-bias conductance maps is shown by the color scale bar. The figure shows (open circles) the obtained intervortex distance vs the magnetic field. The blue line is the prediction for a hexagonal Abrikosov lattice ($d = (4/3)^{1/4} \sqrt{\phi _0 /H}$). {\bf (b)} Main six Bragg peaks of the vortex lattices shown in (a). Black dashed lines are parallel to the steps (shown as white lines in Fig. \ref{Fig1} and Fig. \ref{Fig8} (see appendix)). {\bf (c)} Normalized tunneling conductance $\sigma$ vs. normalized distance $\rho$ obtained from vortices at 0.4 T and 0.8 T (see text for the normalization procedures). The inset shows the vortex core size $\cal C$ vs the magnetic field. Dots are the values of $\cal C$ obtained from the fits given by the lines in the main panel of (c). The continuous black line is given by $\cal C$ $\propto 1/\sqrt{H}$ (see Ref. \cite{Fente16}). {\bf (d)} The magnetic field dependence of the tunneling conductance at 150 mK obtained at the center between vortex cores. Inset shows the magnetic field dependence of the zero bias conductance σ(V = 0 mV). Line is a guide to eye.}
\label{Fig3}
\end{figure*}

Results under magnetic fields are shown in Fig.\ref{Fig3}. We find a hexagonal vortex lattice between 0.2 T up to 2.0 T (H$_{c2}\approx$ $2.5$ T \cite{Udhara16}), with the intervortex distance decreasing with field  as expected. We do not observe signatures of Caroli-de Gennes-Matricon Andreev core states\cite{Caroli64,Hess90,Guillamon08}. Rather, the normalized tunneling conductance at the vortex center reaches one (Fig.\ref{Fig3}c). This is not surprising, taking into account that Rh$_9$In$_4$S$_4$ is strongly in the dirty limit and that the discrete core states are expected to turn into a continuum in the dirty limit\cite{Renner91}.

The vortex lattice is mostly oriented along the directions defined by the steps observed at the sample surface (see Fig. \ref{Fig1} and Fig.\ref{Fig8} at the appendix). Fig.\ref{Fig3}(b) shows the position of the six lowest Bragg peaks of the Fourier transform of the vortex lattice images obtained at different magnetic fields vs the direction of the surface steps. The vortex lattice orients along one of the two crystalline directions of the square atomic lattice at the basal plane. Likely this shows the influcence of large scale defects, such as grain boundaries or large size dislocations, that follow the symmetry of the crystalline structure and pin the vortex lattice.

We have also analized the vortex core radius $\cal C$, which we define as in Refs.\cite{Fente16,arxivKogan}, namely the inverse of the slope of the radial variation of the superconducting order parameter from the vortex center. $\cal C$ can be obtained from the STM images as described in Refs.\cite{Fente16,arxivKogan}. To this end, we make an angular average of the tunneling conductance $\sigma$ (normalized in such a way as to provide unity at the vortex center and zero in between vortices) around the center of a vortex and fit the result to the radial dependence of the density of states discussed in \cite{Fente16,arxivKogan}. The experimental result is shown as points and the fits as lines in Fig.\ref{Fig3}c. We use only images made at two magnetic fields, where we obtained sufficient contrast to perform the analysis. This is admitedly too little to provide a serious magnetic field dependence. Still, we find values for $\cal C$ (inset of Fig.\ref{Fig3}c) that are consistent with the magnetic field dependence proposed in Ref. \cite{Fente16} $\cal C$ $\propto 1/ \sqrt{H}$. We can adjust the $\cal C$ vs H dependence in such a way as to cross the two measured points and obtain at the same time an extrapolation to H$_{c2}$ that provides the same value as the superconducting coherence length obtained from H$_{c2}$, $\xi\approx$ 9.4 nm.

Finally, we have analyzed the intervortex density of states by taking tunneling conductance curves  at the lowest temperatures in the middle between vortices. The result is shown in Fig.\ref{Fig3}d. The intervortex density of states at the Fermi level increases linearly with the magnetic field.

\vspace{-0.4cm}
\section{Vortex creep}
\vspace{-0.3cm}

\begin{figure}[h!]
\includegraphics[width=0.45\textwidth]{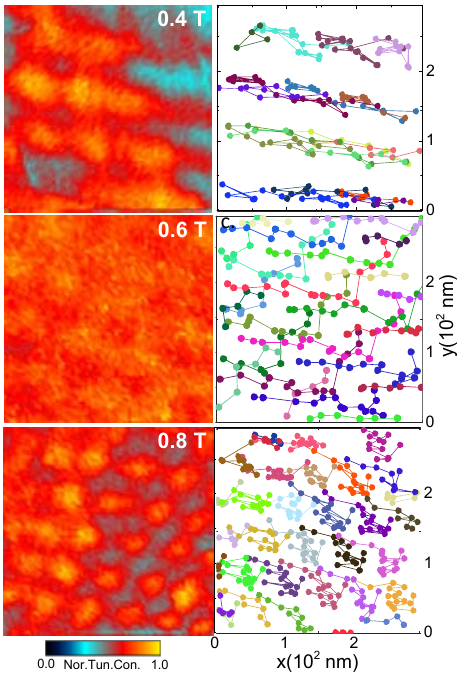}
\vskip -0.2cm
\caption{(Color) In the left panels we show averaged images over all frames of the videos provided as a supplement material \protect\cite{Video}. Some of the frames are given in the Appendix. In the supplemental material we provide a video of moving vortex lattices at 0.15 K and 0.4 T, 0.6 T and 0.8 T respectively. Color scale is given by the bar at the end of the panel. In the right panels we show trajectories of vortices (points). Lines joining the points provide the sequence of motion between consecutive images.}
\label{Fig4}
\end{figure}

We find that vortices are slowly moving during imaging. Images in Fig.\ref{Fig3} were made within a few minutes to obtain static views of the vortex lattice. In Fig.\ref{Fig4} we show experiments made in different locations and magnetic fields. Motion remains unaltered (with similar velocity and direction of motion) at temperatures as low as 150 mK and after more than one day.  We acquired consecutively images taken in a few minutes each during a period of several hours. In Fig.\ref{Fig4} we show the average over a set of consecutive images (left panels of Fig.\ref{Fig4}) and the vortex trajectories during the whole experiment (right panels of Fig.\ref{Fig4}) (see Appendix for a larger set of vortex images at each magnetic field). We include in this figure zero field cooled (upper and lower panels) and field cooled experiments (middle panel). We find that the vortex motion strongly depends on location and magnetic field-temperature history. Generally, we observed either net vortex motion along a well defined direction (upper panel), net motion involving changes in the directions (middle panel) and fluctuations around the vortex positions (lower panel). The average of vortex velocities estimated by considering positions of vortices in subsequent frames is of $100$ \ nm/h in the upper panel, of $40$ nm/h in the middle panel and close to zero in the lower panel.

In the location where we took images of the upper panel, we find that vortices move along a main axis of the vortex lattice and that vortex motion is smooth and correlated, without any signature of jumps or steps. In the location of the middle panel, vortex motion is also along one main axis of the vortex lattice, but vortices move along another axis after a certain time. In the location of the lower panel, we only observe wiggling, without a net motion along a given direction.

\begin{figure*}
\includegraphics[width=1\textwidth]{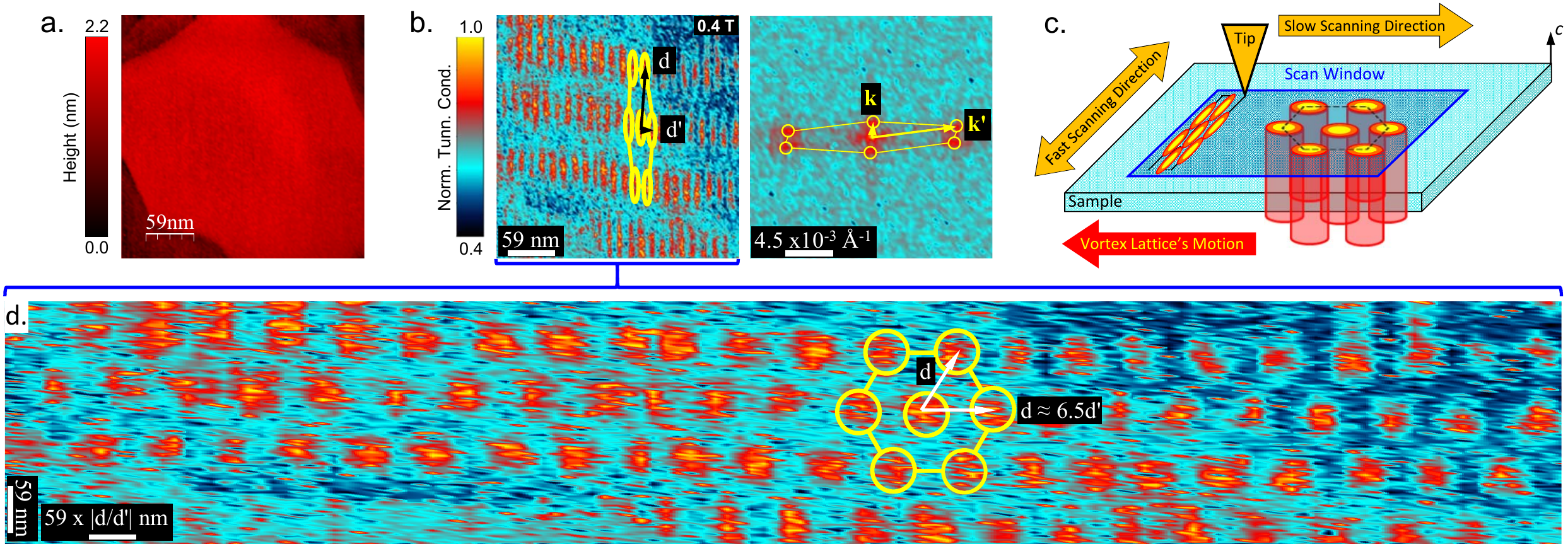}
\vskip -0.2cm
\caption{(Color) (a) Topography image of the spectroscopy map at 0.4 T and 0.15 K provided in left panel of (b). Color scale is given by the vertical bar. (b) {\it (left)} Single image of a moving vortex lattice, taken during 16 hours. Due to vortex motion, the vortex lattice appears squeezed in one direction. We mark one hexagon with yellow circles and the observed lattice constants $d$ and $d'$ with black arrows. From its Fourier transform {\it (right)}, we find that the intervortex distance along the y-axis of the image corresponds to the value expected at 0.4 T, $d \approx 75$ nm. Along the axis where vortices move, we find $d' \approx 11$ nm. (c) A cartoon of the situation found in (b). The vortex lattice (hexagonal cylinders) moves along the direction given by the red arrow. The direction of the scan is given by the orange arrows. The scanning window is schematically marked by a blue square. (d) The same image shown in (b-{\it left}), but stretched along the x-axis by $\frac{d}{d'}$. We show again a single hexagon in yellow.}
\label{Fig5}
\end{figure*}

To show the correlated nature of net vortex motion we have acquired an image  (Fig.\ref{Fig5}) during 16 hours at the same scanning window where we did the experiment shown in the upper panel of Fig.\ref{Fig4}. Scanning was very slow, so that vortices were moving below the tip. The fast scanning axis is along a direction nearly perpendicular to vortex motion. Therefore, vortices appear compressed along the slow scanning direction (see (Fig.\ref{Fig5}b). By stretching the image with the intervortex distance at this magnetic field (0.4 T), we reproduce a hexagonal vortex lattice (Fig.\ref{Fig5}d). To understand this, we must consider the relation between vortex and tip motion. Vortices have to move with a constant velocity over the time frame given by the time a vortex remains below the tip, which is of approximately 12 minutes. The vortex lattice moves in the opposite direction to the slow scanning motion. The tip velocity along this direction is of $v_{tip}= 18$ nm/h. Using this velocity and the ratio between the intervortex distance expected for this field, $d$, and the value found in the compressed vortex lattice, $d'$, we obtain for the vortex velocity $v = (d/d' -1 ) v_{tip} =$  100 nm/h. This value coincides with the value measured by making fast consecutive images discussed above in the same field of view (upper panels of Fig.\ref{Fig4}).

Creep can be thermal and/or quantum activated. The relevant scales are given, respectively, by the collective pinning energy $U_c = H_c^2 \xi^3 (j_c/j_0)^{1/2}$ and the Euclidean action $S_E = (\hbar/Q_u)(j_0/j_c)^{1/2}$ with $H_c$ the thermodynamic critical field, $\xi$ the coherence length, $j_c$ and $j_0$ the critical and depairing current densities and $Q_u$ the quantum resistance \cite{Blatter94}. The crossover temperature $T_0 = \hbar U_c/S_E$ determines the temperature below which quantum effects dominate. In Rh$_9$In$_4$S$_4$, we find $U_c = 3.5 K$ using $\mu_0H_c = 25 mT$, $\xi = 9.4$ nm \cite{Udhara16}, $j_c = 3 \cdot 10^8 A/m^2$ \cite{Hughes01} and $j_0 = 3 \cdot 10^{10} A/m^2$. On the other hand, superconductors with a large normal state resistivity and small coherence lengths, i.e. quantum resistance $Q_u = (e^2/\hbar)(\rho_n/\xi)$ of order of a k$\Omega$ are candidates for quantum creep\cite{Blatter93}. In Rh$_9$In$_4$S$_4$, $Q_u$ is of $4 \cdot 10^{-2}$ in between the values reported for conventional superconductors ($Q_u \sim 10^{-3}$) where quantum effects are negligible and cuprate high $T_c$  superconductors ($Q_u \sim 10^{-1}$) where quantum fluctuations are large\cite{Blatter94,PhysRevLett.76.1529,Eley16}. When comparing thermal and quantum contributions, we find that the crossover temperature $T_0$ is of about $15 mK$  indicating that, even at 150 mK, motion is likely thermally activated.

\vspace{-0.5cm}

\section{Discussion and conclusions}
\vspace{-0.3cm}
In summary, we have imaged the hexagonal superconducting vortex lattice in the extreme type-II s-wave superconductor Rh$_9$In$_4$S$_4$. We determine the superconducting coherence length and find values comparable to those obtained using macroscopic upper critical field measurements. We observe directly vortex creep at very low temperatures, $T/T_c<0.1$ in nearly equilibrium conditions. From our direct observation, we conclude that the creep velocity strongly dependens on each experiment. During creep, vortices move by jumping over the pinning barriers due to thermal excitation. The vortex lattice remains hexagonal over length scales well above the image size, comprising here several hexagons. For these small sized vortex bundles and the small driving currents present in our experiment, pinning of isolated vortices plays a minor role in the motion, which rather reflects collective activated motion of the hexagonal lattice over the pinning landscape.

On the other hand, we should consider the intrinsic disorder of the vortex lattice, characterized by absence of positional ordering at large length scales \cite{Larkin70,Nattermann90,Giamarchi94,Giamarchi95}. The consequence of disorder is expected to manifest in the dynamic properties of the vortex lattice by the fact that the height of the pinning barriers $U$ depends on $j$ for $j<<j_c$. In particular, $U$ was shown to diverge as $U\propto j^{-\alpha}$ with decreasing $j$\cite{Feigelman89}. This actually leads to a truly superconducting (dissipationless) phase also in systems with large vortex creep\cite{Blatter94,Brandt95}. In our experiments we observe very different dynamic behavior for configurations in which the current producing vortex motion $j<<j_c$, which depends on the particular pinning landscape at each location, has been certainly very different too. The fact that we observe moving as well as nearly static vortex lattices shows that the barrier height  $U$ for vortex motion strongly depends on location and magnetic field-temperature history.

\begin{figure}[h!]
\includegraphics[width=0.45\textwidth]{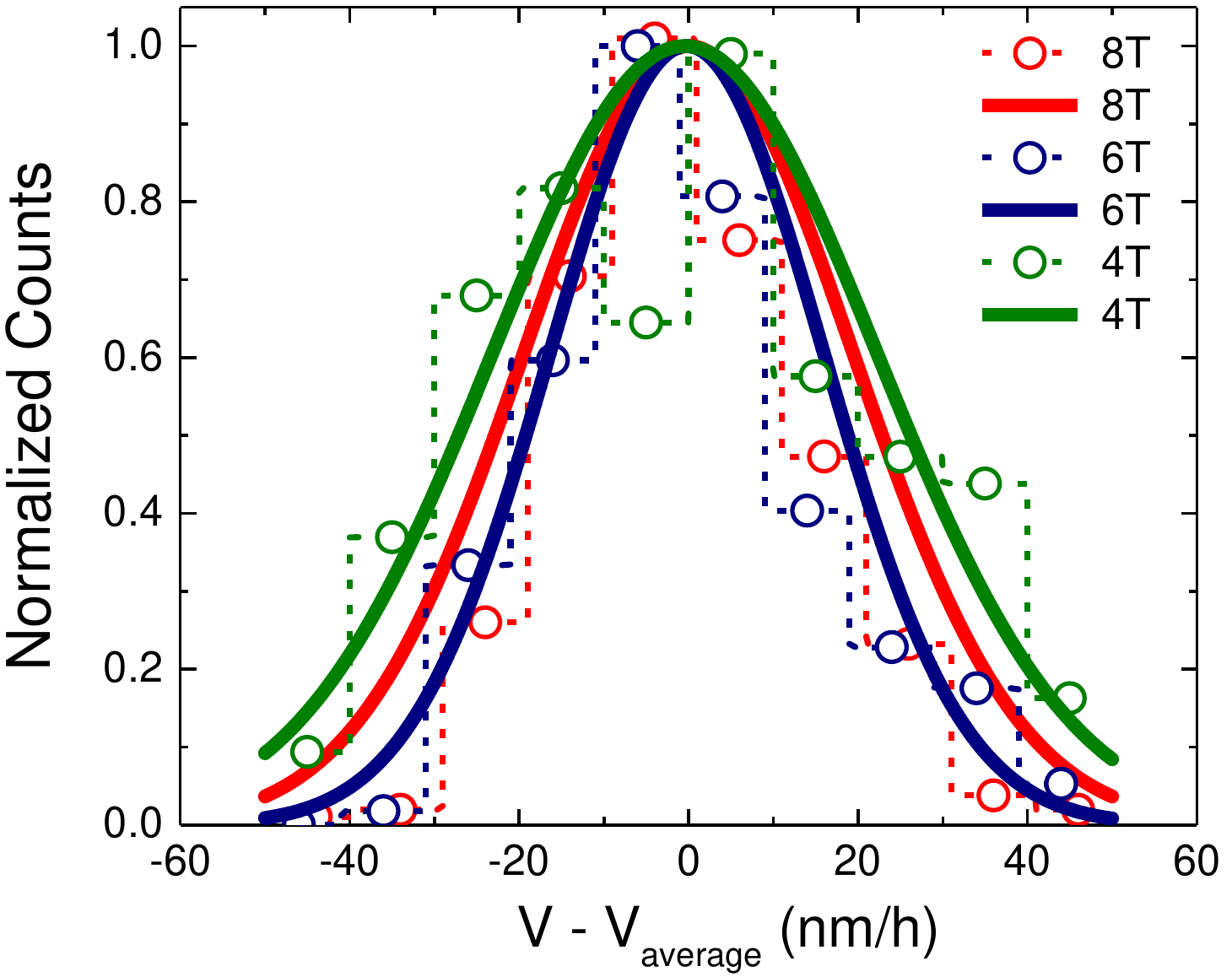}
\vskip -0.4cm
\caption{(Color online) Histogram of the velocities found in each experiment discussed in Fig.\ref{Fig4}. We represent counts for a given velocity (x-axis) measured with respect to the average velocity. The counts (y-axis) are normalized to one at the center. To calculate the velocity, we use the distance travelled between two subsequent images. The histograms are shown by the dashed lines and points and the lines are Gaussian fits, giving approximately a half width of 23 nm/h, 16 nm/h and 20 nm/h for 0.4 T, 0.6 T and 0.8 T respectively.}
\label{Fig5a}
\end{figure}

Interestingly, we observe that the velocity fluctuates significantly around the average values provided above (see Fig.\ref{Fig5a}). If we follow the velocities of all vortices as a function of time, we observe that all motion is accompanied by wiggling vortices, resulting in a roughly Gaussian distribution of velocities with a width between 16 nm/h and 23 nm/h. The resulting fluctuations in the vortex positions are somewhat smaller but of order of the intervortex distance (as can be readily seen in the Fig.\ref{Fig4}). Instead of moving smoothly, the vortex lattice makes jumps trying adjacent metastable configurations, showing that the pinning potential has many small minima that are superimposed to a large variation producing creep\cite{Blatter94}. As the barrier for motion is large for small currents, the wiggles just provide intermediate and unstable arrangements.

So our images of creeping vortex lattices visually show the capability of the vortex lattice to deform and adapt to the pinning landscape, giving glassy dynamic properties while maintaining hexagonal order.

\vspace{-0.4cm}
\section{Acknowlegments}
\vspace{-0.3cm}
We acknowledge R. Willa and V.G. Kogan for helpful and critical comments and discussions with S. Vieira. We wish to acknowledge the support of Departamento Administrativo de Ciencia, Tecnolog\'ia e Innovaci\'on, COLCIENCIAS (Colombia) Programa Doctorados en el Exterior Convocatoria 568-2012. We further acknowledge support by the Spanish Ministry of Economy and Competitiveness (FIS2014-54498-R, MDM-2014-0377), by the Comunidad de Madrid through program Nanofrontmag-CM (S2013/MIT-2850), by EU (IG, European Research Council PNICTEYES grant agreement 679080, FP7-PEOPLE-2013-CIG 618321, Cost MP-1201) and by Axa Research Fund. SEGAINVEX-UAM and Banco Santander are also acknowledged. Work done in Ames Lab was supported by the U.S. Department of Energy, Office of Basic Energy Science, Division of Materials Sciences and Engineering. Ames Laboratory is operated for the U.S. Department of Energy by Iowa State University under Contract No. DE-AC02-07CH11358.

\vspace{-0.5cm}
\section{Appendix}
\vspace{-0.6cm}
\begin{figure}[h!]
\includegraphics[width=0.45\textwidth]{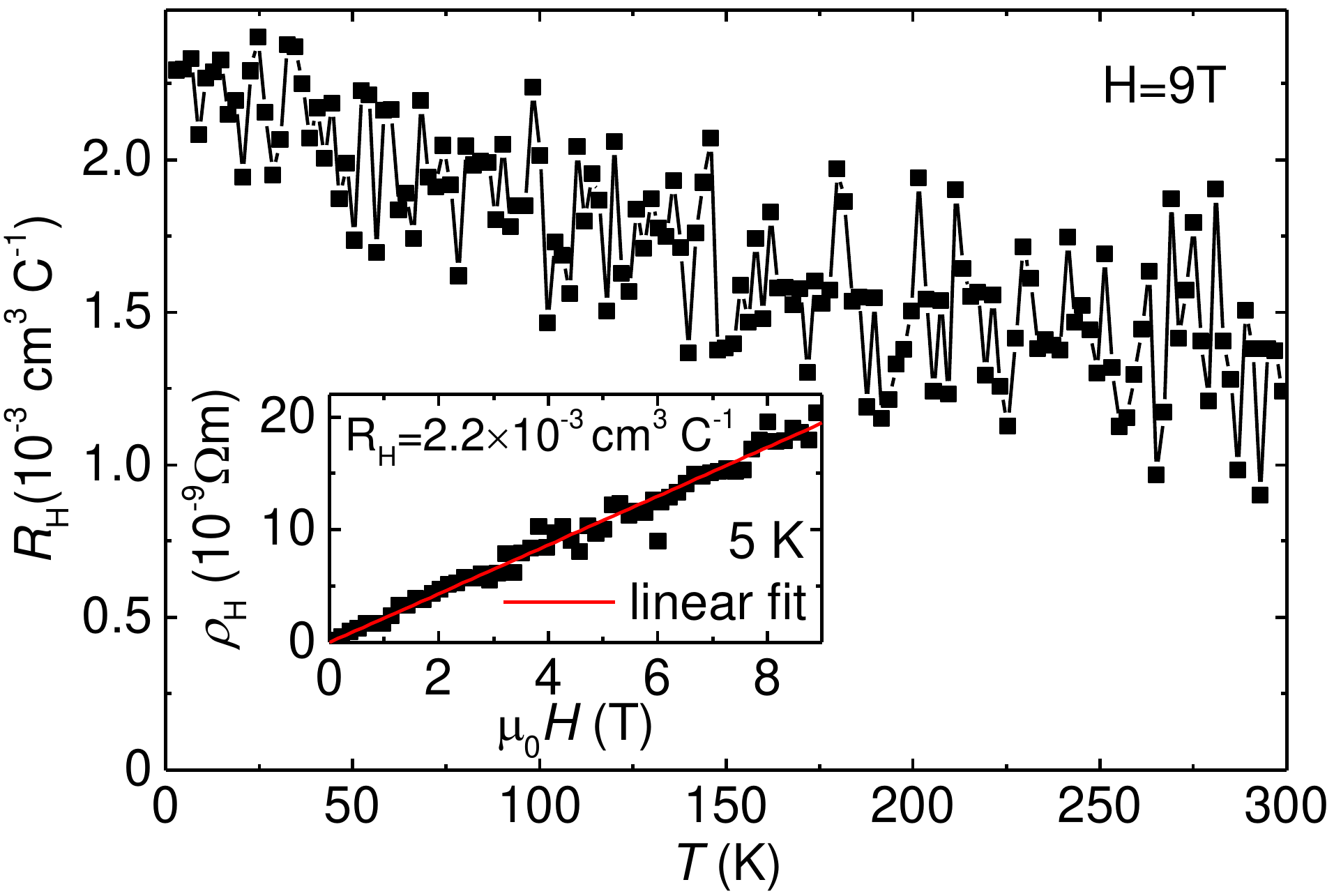}
\caption{Hall coefficient at 9 T as a function of temperature. The inset shows the Hall resistivity at 5 K. The red solid line is the fit using a single band model with a carrier density of $2.8 \times 10^{21}$ cm$^{-3}$ .}
\label{Fig6}
\end{figure}
\begin{figure*}[h!]
\includegraphics[width=0.9\textwidth]{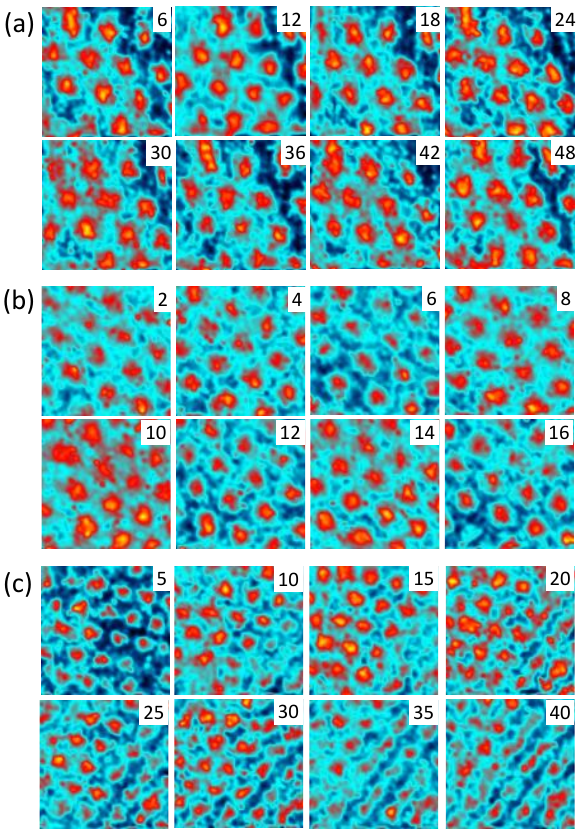}
\caption{(Color online) Some of the tunneling conductance maps of the moving vortex lattice in Rh$_9$In$_4$S$_4$ shown inf Fig. \ref{Fig4} at 0.4 T (a), 0.6 T (b) and 0.8 T (c) at 0.15 K. ($\sim$18 min/map). Numbers indicate the frame at each magnetic field. Same color scale as in Fig. 4.
}
\label{Fig7}
\end{figure*}
\begin{figure*}[h!]
\includegraphics[width=0.7\textwidth]{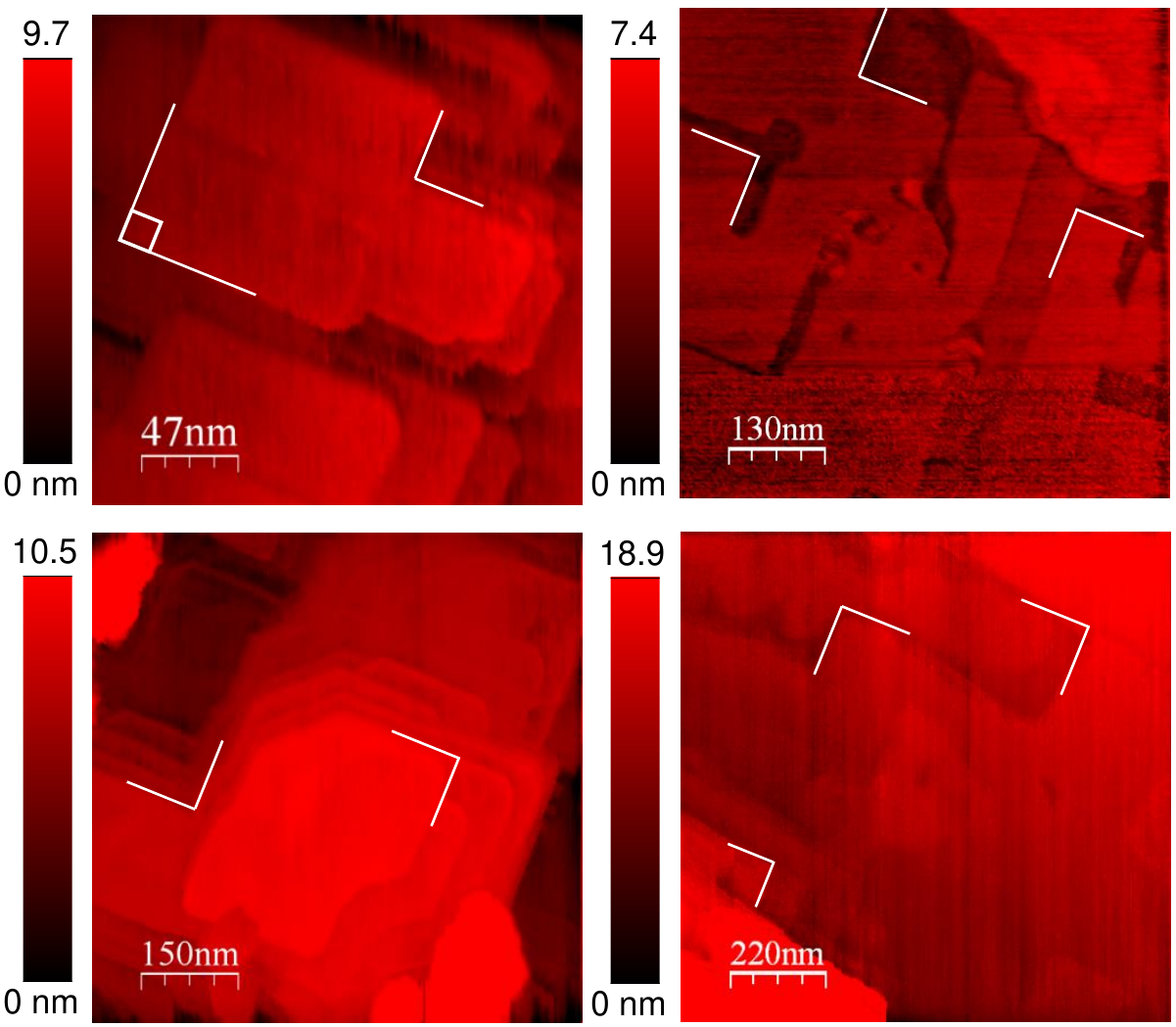}
\caption{(Color online) Typical topographic images obtained in Rh$_9$In$_4$S$_4$. All are taken at a bias voltage of 5 mV and a tunneling current of 0.2 nA. The images show characteristic square shape features, with step height separating these of order of an integer of the unit cell.}
\label{Fig8}
\end{figure*}

\vspace{-0.4cm}
Figure \ref{Fig6} shows the Hall response of the single crystals of Rh$_9$In$_4$S$_4$. Calculated carrier density using single band model  at 5 K is $2.8 \times 10^{21}$ cm$^{-3}$. Using this value for the carrier density, the Drude model and the resistivity \cite{Udhara16} we estimate a mean free path of $\ell\approx 5$ nm. In Figure \ref{Fig7} we show a set of the frames from the experiment shown in Fig.\ref{Fig4}. In Fig.\ref{Fig8} we show topographic images obtained in several places of the sample.

\end{document}